\documentclass[aps,preprint,a4paper,12pt]{revtex4-1}
\usepackage[utf8]{inputenc}
\usepackage{graphicx,amssymb,amsmath,amsthm,amsfonts,amscd, mathrsfs,epsfig,epsf}
\usepackage{hyperref}
\usepackage{amsfonts}
\usepackage{amssymb}
\usepackage{graphicx}
\usepackage{latexsym}
\usepackage{color}
\usepackage{booktabs}
\usepackage{dcolumn}
\usepackage{epsfig}
\usepackage{subfigure}
\usepackage{float}
\usepackage{multirow}
\usepackage{ulem}
\usepackage{xcolor}
\usepackage{svg}
\usepackage[mathcal]{euscript}
\usepackage{tikz}
\def\be{\begin{eqnarray}}
\def\ee{\end{eqnarray}}

\begin{document}

\title{Scalar field instabilities in charged BTZ black holes}

\author{R. D. B. Fontana}
\email{rodrigo.dalbosco@ufrgs.br}
\affiliation{Universidade Federal do Rio Grande do Sul, Campus Tramandaí-RS
\\
Estrada Tramandaí-Osório, CEP 95590-000, RS, Brazil \\
Departamento de Matemática, Universidade de Aveiro and \\
Center for Research and Development in Mathematics and Applications (CIDMA), Campus Santiago, 3810-183, Aveiro, Portugal}


\date{\today}

\begin{abstract}

We investigate the charged scalar field propagating in a (2+1) charged BTZ black hole. The conditions for stability are studied unveiling, for each black hole geometry, the existence of a critical scalar charge as of the evolution is unstable. The existence of growing profiles is substantiated by the depth in the effective potential that intensifies as the scalar charge increases. The phenomenon happens in every black hole geometry even for small geometry charge. In the small scalar charge regime, the field evolution is stable and in such we calculate the quasinormal modes.

\end{abstract}



\maketitle

\section{Introduction}\label{sec1}

Lower dimensional gravity is an active field of research in the last decades in many different directions. Some of the pioneer works were launched in the 1980s and 1890s \cite{jackiw1, dese1, dese2, brown1, brown2, mana1, mann1995lower} and a decade later singular solutions were discovered \cite{Banados:1992wn, Carlip_1995, Martinez_2000} describing spacetimes with mass, charge, rotation and cosmological constant. Although no graviton is to be found in such theory (since no dynamical degrees of freedom are present) the nontrivial spacetime solutions of the curvature equations produce valuable dynamical consequences worth of investigation and simpler as the the 4-dimensional counterpart theory (for a detailed study of the different geometries in (2+1)-dimensions, refer to \cite{Podolsk__2022}). 

In particular the black holes of lower dimensional gravity exhibit interesting properties simpler then that of general relativity. As an example, demonstrated in \cite{Birmingham_2002}, the poles of the retarded correlation function of the 2-dimensional conformal field theory represent exactly the quasinormal modes of the solution, emphasizing the interpretation of their imaginary part as relaxation time in the perturbed regime. 

In the present work we will be concerned with the statical BTZ black hole with charge. From the perspective of thermodynamics, the Wald semiclassical limit for particles absorption was tested in \cite{Song_2018}, the charged case in \cite{Gwak_2016} and other semiclassical aspects considered in \cite{Liu_2006, Jiang_2007, Li_2007, larranaga2007law, Ejaz_2013}. The geodesic motion of test particles in the nonlinear regime was examined in \cite{Soroushfar_2016, Panotopoulos_2022},  the scale dependence in \cite{Rincon_2017, Panotopoulos_2018, Rincon_2018} and the accelerated version of the black hole in \cite{Astorino_2011}. Other solutions of (2+1) dimensional charged black holes in alternative theories or accelerated geometries are found in \cite{Panah_2018, Eslam_Panah_2023, panah2022threedimensional}.

In this paper we will be interested particularly in a charged scalar field that propagates in a charged BTZ geometry. The scalar perturbation considering a nonlinear Maxwell term was studied e. g. in \cite{Gonzalez_2021, Aragon_2021} (with/without scalar charge) and in \cite{fon23} in the linear theory. 

Charged scalar perturbations and instabilities in charged black holes are expected to coexist, in strict relation to the superradiance of real waves \cite{kono1, Destounis_2019, wang14, Zhu_2014}. The superradiance in BTZ geometries was analyzed in different studies, e. g. with rotation in \cite{Dappiaggi_2018, Ortiz_2012, Ho_1998} and in \cite{kone} with the Maxwell charge, but not considering a perturbative treatment. In AdS and asymptotically flat charged black holes a thorough investigation is performed in \cite{Dias_2012,Dias_2017,Dias_2018}.

We will study the charged scalar field perturbation in the BTZ geometry, focusing on the instabilities that may be present. In the next section, we introduce the fundamentals of this black hole and the numerics used to evolve the scalar field in the background, following with the instability analysis in section \ref{sec3}. In section \ref{sec4} we calculate the quasinormal modes (QNMs - for an incomplete list with other studies on QNMs in BTZ geometries, refer to \cite{Cardoso_2001, Fernando_2004, Crisostomo_2004, LO_2006, Fernando_2008, Fernando_2009, Kwon_2010, Gonz3, Rinc3, Pano4, rod2, pano5, Konoplya_2020, Aragon_2021, Cuadros_Melgar_2022, fon23, de_Oliveira_2024}) and unstable frequencies for different scalar charges presenting our conclusions in section \ref{sec5}.

\section{Charged BTZ black hole and numerics}\label{sec2}

We start by considering a charged BTZ black hole without rotation with metric
\be
\label{e1}
ds^2= -fdt^2 + f^{-1}dr^2 + r^2 dx^2,
\ee
in which
\be
\label{e2}
f= -M + r^2 - Q^2 \ln (r)
\ee
is the lapse function. The geometry (\ref{e1}) is spherically symmetric since we are considering $x$ as an angular variable with a $2\pi$ range. Such choice brings nontrivial consequences in stablishing the ansatz for the scalar field (as described below).
The spacetime has two horizons (Cauchy  - $r_c$ - and event - $r_h$). For each specific value of $r_h$, the maximum value of the charge reads $Q_{max}= \sqrt{2r_h}$ which is not extremal \cite{fon23}. Extremal charged BTZ black holes happen only for $M<1$ divided in two branches according to the path for a naked singularity (relative to the addition or subtraction of charge) \cite{Carlip_1995,Martinez_2000}. 

The lapse functions has its minimum value at $r_{min}=Q/\sqrt{2}$ in which 
\be
\label{e3}
f_{min}=-M+Q^2\left( \frac{1}{2}-\ln \left(\frac{Q}{\sqrt{2}}\right) \right).
\ee
Whenever $f_{min} \leq 0$ we have a spacetime with two horizons covering the singularity at $r=0$ and cosmic censorship (weak) is preserved. We shall take this condition as granted by imposing $M \geq 1$. For our study, it is convenient to define the quantity  $\mathcal{Q}=Q^2/Q_{max}^2 = Q^2/2r_h$ in the range $[0,1]$, that we use to analyze the scalar field dynamics in the next sections.

We want to study the free charged scalar field in the black hole geometry, whose motion equation is written as
\be
\label{e4}
g^{\mu \nu}(\nabla_\mu - iqA_\mu )(\nabla_\nu - iqA_\nu )\Phi_s =0
\ee
in which $A = A_\mu dx^\mu = A_t dt = -Q\sqrt{2}\ln (r) dt$ is the vector potential and $\Phi_s$ the scalar field. Choosing an usual field transformation \cite{fon23}, $\Phi_s \rightarrow e^{-ikx}\frac{\psi}{\sqrt{r}}$ we obtain the Klein-Gordon equation as 
\be
\label{e5}
-\frac{\partial^2}{\partial t^2}\psi + \frac{\partial^2}{\partial r_*^2}\psi + V(r) \psi + 2iqA_t\frac{\partial}{\partial t} \psi + q^2A_t^2\psi =0,
\ee
in which $V(r)$ the scalar potential,
\be
\label{e6}
V(r) = f \left( \frac{f}{4r^2}-\frac{\partial_rf}{2r} - \frac{k^2}{r^2} \right),
\ee
and $dr_*=f^{-1}dr$ is the tortoise coordinate. The ansatz in the $x$ coordinate, $\Phi_s (x) = e^{-ikx}$ fixes the quantization rule $k \in \mathbf{Z}$, settling the continuity of the field and its derivatives at the limits of $x$. We apply an integration scheme in double null coordinates similar to that described in \cite{Mo_2018} to achieve the quasinormal signal with initial data
\be
\label{e7}
\Psi_{r\rightarrow \infty} = 0, \hspace{1.0cm} \Psi_{u_0,v} = Gaussian
\ee
and filter the field profile to obtain the frequencies with the Prony technique \cite{Konoplya:2011qq}. As a second method to check the data we obtain, we use the Frobenius expansion similar to \cite{hor2000}, considering the Klein-Gordon equation in the null direction $v$ instead of $t$. Taking the radial vector $z=1/r$, the scalar equation in such coordinate system reads
\be
\label{e8}
\left( fz^4 \frac{\partial^2 }{\partial z^2} + (2fz^3-z^2\partial_r f + 2i\omega z^2) \frac{\partial}{ \partial z} + \left( \frac{fz^2}{4} + \frac{2q \omega A_t}{f} - \frac{z\partial_r f}{2}-k^2z^2\right) \right) \Psi =0,
\ee
in which $\omega = \Re (\omega )+i\Im (\omega )$ is the eigenvalue to be numerically obtained through the implementation of (\ref{e7}). As in \cite{Mo_2018} we apply a more suitable vector potential $A$ in order to better converge the Frobenius series by choosing $A_t = -Q\sqrt{2}\ln (r/r_h)$, which allows us to avoid the first term of the potential (expanded around the event horizon, $z_+$) of the above equation.

\section{Instabilities}\label{sec3}

In view of previous literature of charged scalar fields propagating in charged spherically symmetric black holes \cite{ Gonzalez_2021, Aragon_2021, kono1, Destounis_2019}, we apprehend two conditions for the presence of field instabilities.

The first is related the behavior of $\Re (\omega )$ (usually referred as the superradiant condition) \cite{kono1}. In the case of the CBTZ black hole, the necessary (but not sufficient) condition for the presence of unstable quasinormal modes is (see appendix \ref{app1} for details)
\be
\label{e9}
\Re (\omega ) > \Phi_h
\ee
where $ \Phi_h$ stands for the value of the electric potential $ \Phi = qQ \sqrt{2}\ln (r)$ at the black hole event horizon. For the particular case where $r_h \leq 1$ every mode fulfill such condition. Even in such case, a second condition must be accomplished as we describe bellow, preventing the $\Phi\rightarrow 0$ situation. Although for small black holes the requirement is trivially satisfied, that is not the case for $r_h >1$. It is worth to mention that the result come as a novelty when compared with similar geometries: in the Reissner-Nordstr\"om black hole \cite{beke1,naka1,crispi1} or the Reissner-Nordstr\"om with anisotropic fluid spacetime \cite{oli21, kono1} the presence of instabilities is limited by a maximum value of $\Phi_h$ for the real part of $\omega$. The reason for such discrepancy is that, in both cases (also in RNAdS geometry) the electromagnetic potential is a monotonically decreasing function, contrary to our spacetime. A figure representing that special behavior is given in panel \ref{fig2}. A similar condition can be obtained when studying the scattering of real-frequency waves in the geometry with physical boundary conditions. Following the usual method of analysis, a superradiant condition can be demonstrated to be given by $\omega > \Phi_h$, consonant with (\ref{e9}).

The second condition is related to the signal of the effective potential in the region between the event horizon and AdS infinity \cite{hor2000, Gonzalez_2021}. If
\be
\label{e10}
\vartheta \equiv V(r) -q^2A_t^2>0
\ee
all modes have $\Im (\omega ) < 0$ being stable (see appendix \ref{app1} for specifics). On the other hand, if we have $V(r) -q^2A_t^2<0$ at last in some region between $r_h$ and AdS infinity, $\Im (\omega ) > 0$ is allowed. 

\begin{figure}
\begin{center}
\includegraphics[width=0.45\textwidth]{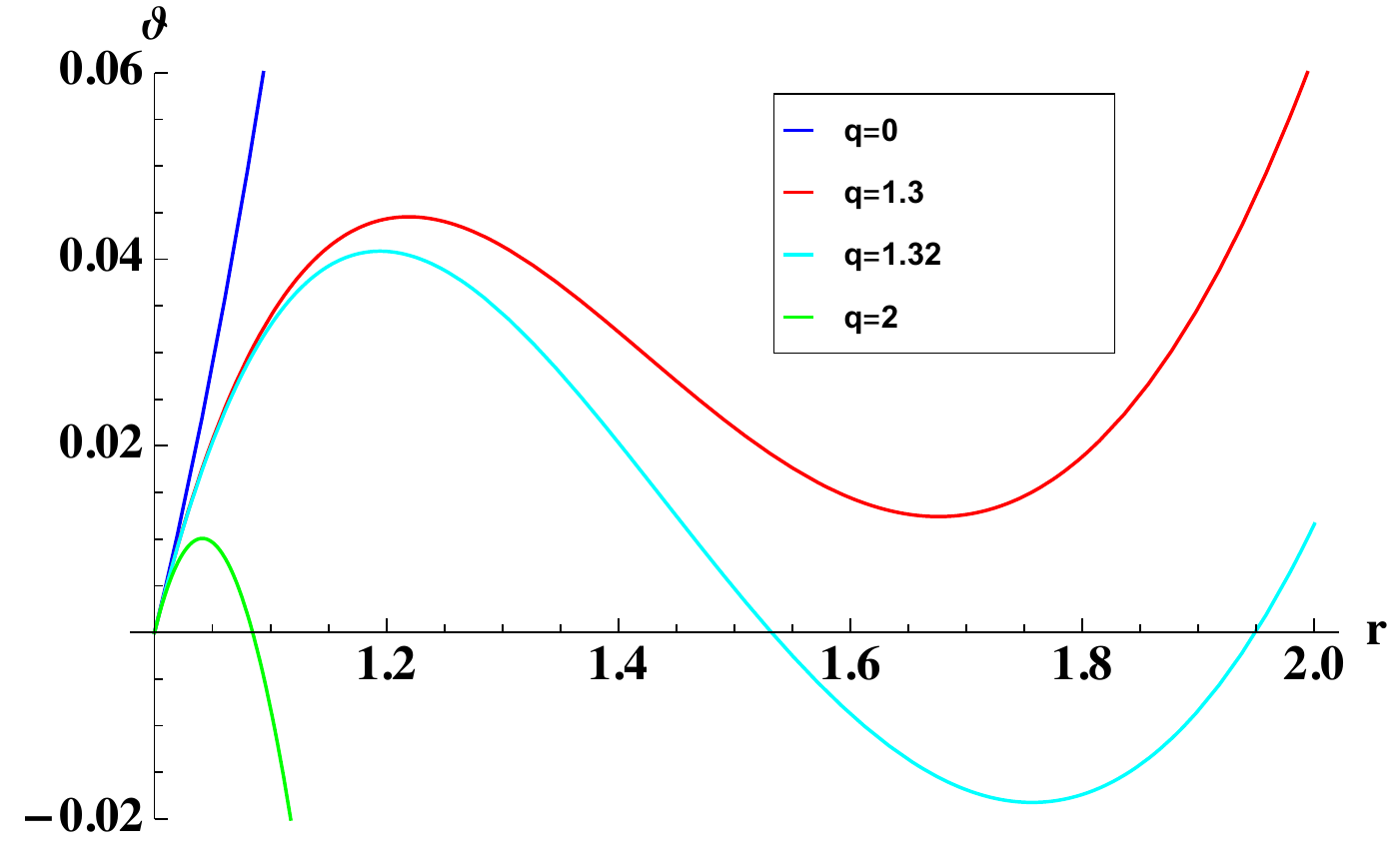}
\includegraphics[width=0.45\textwidth]{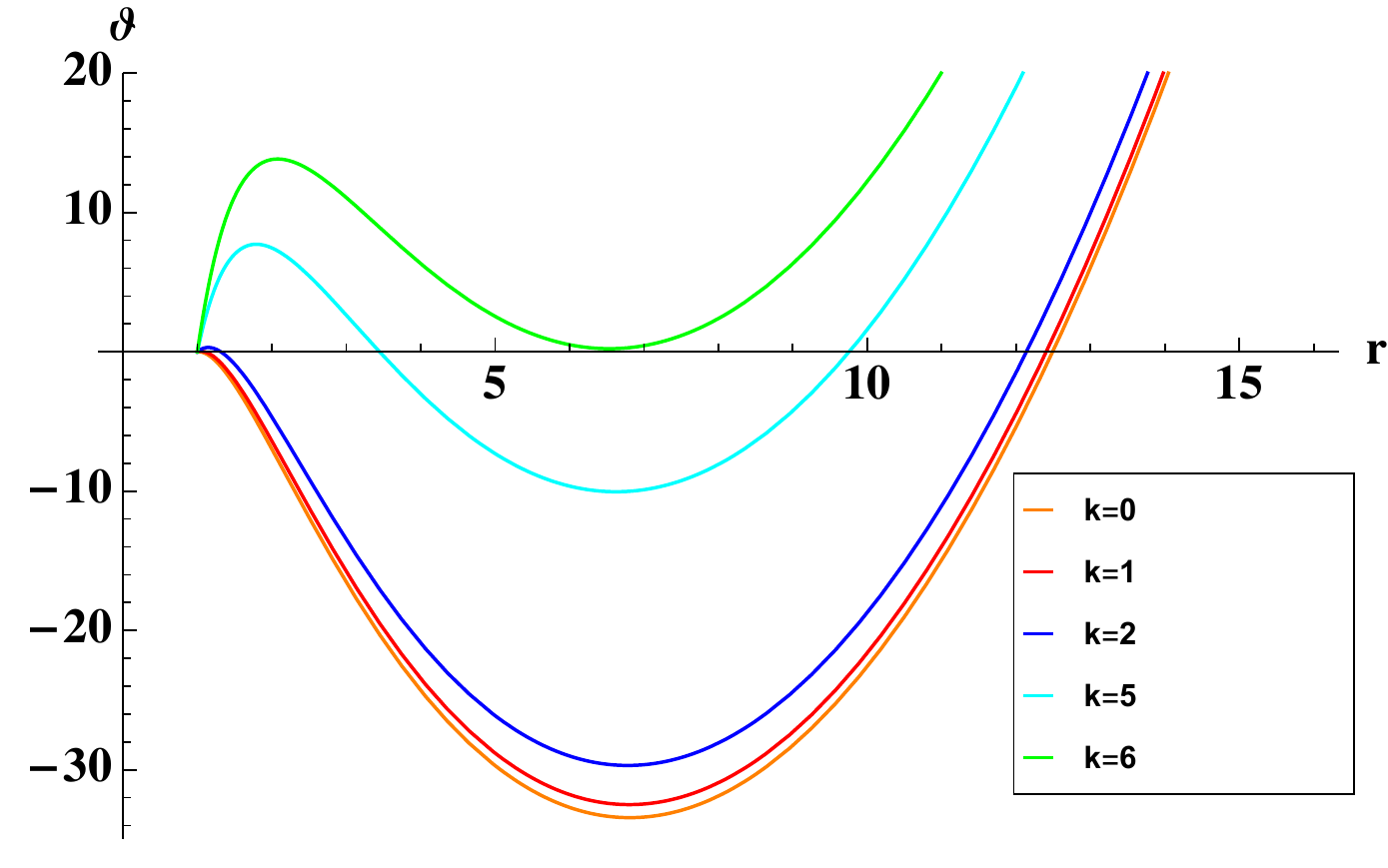}
\end{center}
\caption{The scalar field potential in charged BTZ black hole geometries. The geometry parameters read $r_h=Q=1$ and the scalar parameters $k=0$ (left panel) and $q=3$ (right panel).}
\label{fig1}
\end{figure}

This is the case for sufficient large scalar charges expression as it can be seen in figure \ref{fig1}. As of some particular value of $q$, the potential develops a depth that eventually becomes negative for larger $q$'s. The depth grows indefinitely with increasing charge, implying the existence of a critical scalar field charge $q_c$ for every charged black hole from which point the field is considered unstable ($q>q_c$). It is also worth mentioning that angular momenta $k>0$ bring instabilities as well, for sufficient high scalar charge (see e. g. the right panel of \ref{fig1}), although $k=0$ represents the most unstable case or equivalently the smallest $q$ for stable evolutions.

\begin{figure}
\begin{center}
\includegraphics[width=0.45\textwidth]{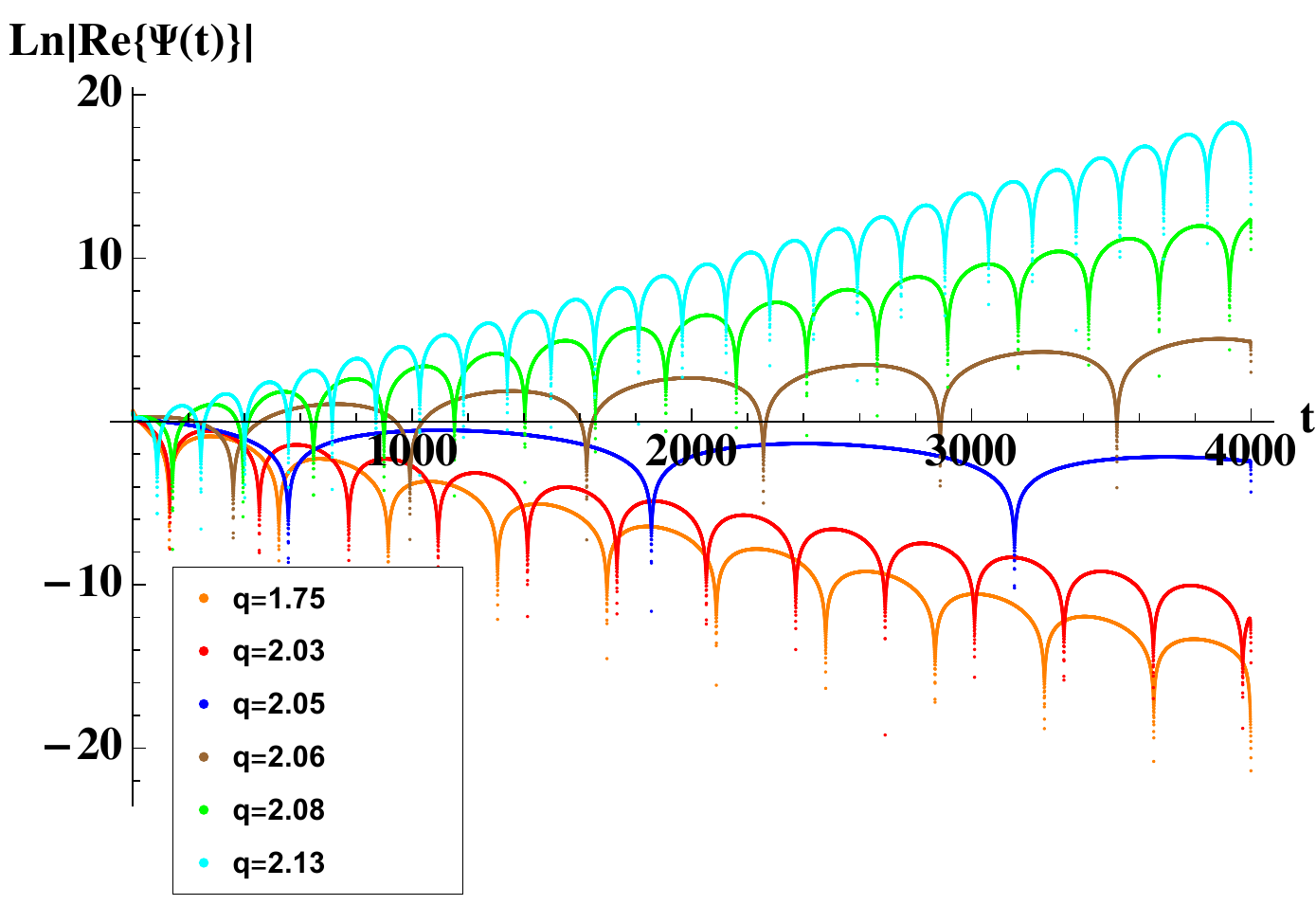}
\includegraphics[width=0.45\textwidth]{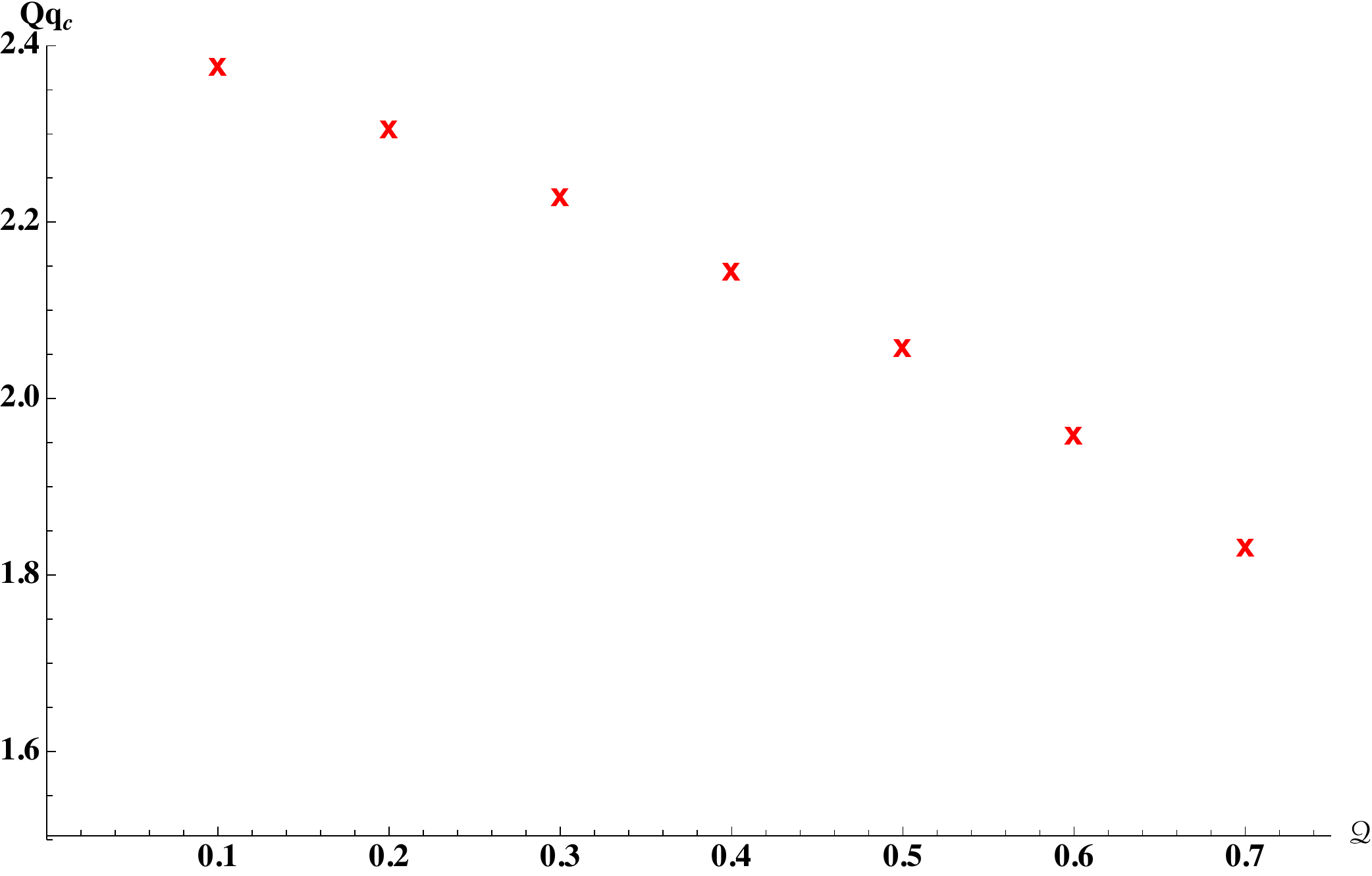}
\end{center}
\caption{The scalar field profiles and critical charges for different BTZ black hole geometries. The geometry event horizon reads $r_h=1$ with scalar angular momentum $k=0$ (in left panel $\mathcal{Q}=0.5$).}
\label{fig2}
\end{figure}

Integrating the differential equation (\ref{e5}) with the method described in \cite{Gundlach_1994} (for a specific description see e. g. \cite{fon23}), we obtain the field evolution that presents a threshold for stability for each black hole charge. In figure \ref{fig2}, left panel, we see the critical value $q_c \simeq 2.05$ of the scalar charge for a black hole with $r_h=2\mathcal{Q}=1$ considering field evolutions with different $q_c$'s and $k=0$. Such behavior is general and was found for every black hole investigated considering different $\mathcal{Q}$'s. The presence of a threshold of stability is robust in charged BTZ black holes: depending on the hole $Q$, the frontier for stable profiles is higher or smaller in $q$, but instabilities are always present for sufficient high scalar charge. In figure \ref{fig2} we can see this behavior illustrated: in the left panel the scalar field profiles are stable if $q<q_c \sim 2.05$ and unstable for $q>q_c$. In that case, the higher the scalar charge, the higher the instability (see the different frequencies for different $q$'s in the next section).

In figure \ref{fig2}, the right panel displays the threshold of stability relating the scalar potential term $Qq$ and the absolute charge of the geometry. The panel shows that the higher the geometric charge, the smaller the scalar charge that destabilizes the geometry. Scalar field configurations with $q<q_c$ evolve as damped oscillations, the quasinormal modes. Profiles with higher charges, $q>q_c$ interact nontrivially with the geometry, and are expected to condensate in scalar clouds around the black hole, changing the background spacetime \cite{Brito:2014wla, Brito_2020, Dias1}. That interaction act as a mechanism for the black hole to lose its charge, until the instability settles down if the scalar charge is not high.

The timescale for such classical instability is not bounded by the usual limitations of charged spacetimes. That is, the instability is enhanced as the scalar charge increases and this is not the case for specific charged spacetimes in four dimensions. In these geometries, the limitation on $\Re (\omega )$ provided by $\Phi_h$ prevents the timescale of instability $\tau_i$ to be of order of the scalar cloud oscillation period ($\tau_o \sim 1/ \Re (\omega )$). As a consequence, $\tau_i \gg \tau_o$ and the instabilities are mild (very small $\Im (\omega )$ e. g.  \cite{oli21, kono1}). In such sense they coexist with the event horizon extracting charge of it until balance of the background is restored. 

In the absence of a constraining mechanism, the response on the final stage of the geometry considering the triggered instability for high scalar charges is outside the scope of this work and has to be investigated with the full nonlinear perturbation theory.

The scalar field configurations with small charges evolve stably, decaying exponentially through the quasinormal modes. We calculate those frequencies (or the unstable growing mode when it is the case) using the Prony technique in the next section.
\section{Quasinormal frequencies and unstable modes}\label{sec4}

\begin{table}[h]
\begin{center}
  \centering
 \caption{{\color{black} Quasinormal modes and unstable frequencies of the massless charged scalar field without angular momentum ($k=0$) for a charged BTZ black hole with $r_h=1$. The positive values listed are stable,  $\Im (\omega ) < 0$}.}
\addtolength\tabcolsep{4pt}
    {\color{black}\begin{tabular}{c|cccccccc}
    \hline    \hline
	 &	\multicolumn{2}{c}{$\mathcal{Q}=0.1$}  &	\multicolumn{2}{c}{$\mathcal{Q}=0.3$}  &	\multicolumn{2}{c}{$\mathcal{Q}=0.5$}  &	\multicolumn{2}{c}{$\mathcal{Q}=0.7$} \\
$q$ & $\Re (\omega )$ & $-\Im (\omega)$ & $\Re (\omega )$ & $-\Im (\omega)$ & $\Re (\omega )$ & $-\Im (\omega)$ & $\Re (\omega )$ & $-\Im (\omega)$  \\
\hline \hline
0.1	&	0.076176	&	1.439703	&	0.060603	&	0.961214	&	0.043370	&	0.618032	&	0.026291	&	0.339187	\\
0.2	&	0.147961	&	1.421744	&	0.119667	&	0.94883		&	0.085908	&	0.608882	&	0.052119	&	0.333115	\\
0.3	&	0.212973	&	1.395426	&	0.175797	&	0.92884		&	0.126764	&	0.593759	&	0.076993	&	0.322935	\\
0.4	&	0.270768	&	1.363693	&	0.227847	&	0.902112	&	0.165060	&	0.572897	&	0.100311	&	0.308604	\\
0.5	&	0.321839	&	1.328619	&	0.274983	&	0.869688	&	0.199919	&	0.546676	&	0.121355	&	0.290142	\\
0.6	&	0.366935	&	1.291527	&	0.316684	&	0.832659	&	0.230508	&	0.515647	&	0.139287	&	0.267692	\\
0.7	&	0.406781	&	1.253241	&	0.352699	&	0.792073	&	0.256104	&	0.480526	&	0.153182	&	0.241604	\\
0.8	&	0.442002	&	1.214289	&	0.382982	&	0.748872	&	0.276136	&	0.442144	&	0.162107	&	0.212473	\\
0.9	&	0.473116	&	1.175015	&	0.407629	&	0.703874	&	0.290213	&	0.401397	&	0.165207	&	0.181139	\\
1	&	0.500549	&	1.135654	&	0.426819	&	0.657766	&	0.298112	&	0.359177	&	0.161767	&	0.148635	\\
2	&	0.628472	&	0.759345	&	0.367113	&	0.225175	&	0.037119	&	0.010730	&	0.366647	&	-0.035514	\\
3	&	0.578287	&	0.439879	&	0.070318	&	-0.014484	&	0.943204	&	-0.061985	&	1.894894	&	-0.079005	\\
4	&	0.404143	&	0.193551	&	0.940762	&	-0.067772	&	2.524917	&	-0.099083	&	4.070260	&	-0.121084	\\
5	&	0.118868	&	0.032758	&	2.167639	&	-0.097798	&	4.506017	&	-0.134793	&	6.706323	&	-0.162825	\\
6	&	0.295264	&	-0.042065	&	3.644955	&	-0.126166	&	6.794021	&	-0.170221	&	9.702066	&	-0.205235	\\
7	&	0.855631	&	-0.070071	&	5.321460	&	-0.153900	&	9.332879	&	-0.205998	&	12.99489	&	-0.249342	\\
8	&	1.537081	&	-0.088964	&	7.164914	&	-0.181437	&	12.08530	&	-0.242819	&	16.54344	&	-0.296610	\\
9	&	2.313147	&	-0.106378	&	9.152723	&	-0.209198	&	15.02505	&	-0.281516	&	20.31969	&	-0.349008	\\
10	&	3.168653	&	-0.123160	&	11.26827	&	-0.237487	&	18.13329	&	-0.323188	&	24.30481	&	-0.409207	\\
 \hline  
    \end{tabular}}
  \label{tb1}
\end{center}
\end{table}

The quasinormal frequencies and unstable modes of the massless charged scalar field are displayed in table \ref{tb1}. The results were obtained through characteristic integration and the Prony methods. We also explore the Frobenius expansion in order to check those results for the quasinormal frequencies.That can be achieved starting from Eq. (\ref{e8}) considering just the same relations as that of \cite{fon23}, Eqs. (13) to (18), adding the charge term to the potential, $u(z)_c=2 f \omega \Phi$. The implementation is similar to that described in \cite{fon23} with this small modification.

The expansion is poorly convergent for high values of charge (scalar, geometric) and small $r_h$. The results with the method are though, less than $0.2\%$ deviant from that of table \ref{tb1}, for $q=0.1$. When $q$ increases, the deviation increases as well to a (maximal of a) few percents for the quasinormal modes with $q \leq 1$. The smaller the $q$ the smaller the deviation between both methods. It is though not possible to calculate unstable frequencies through the Frobenius series as they do not converge in that limit. 

Our results of table \ref{tb1} and figures \ref{fig2} are consonant with that obtained in the nonlinear electrodynamical regime: the higher the black hole charge, the smaller the scalar charge at which instabilities are found. This can be seen e. g. in \cite{Gonzalez_2021}, where unstable modes are present with smaller scalar charges when the geometric charge is increased (see e. g. Tables V and VI).  

In the quasi-adiabatic regime, $q \sim q_c$ when the timescale of the scalar field is large (its energy being small compared to the black hole), we can calculate the amplification factor from the ingoing to the outgoing fluxes, as evidence of the presence of the instabilities through the current,
\be
\label{ee2}
\mathcal{J} = \frac{\sqrt{-g}g^{rr}}{2i}(\Psi^*\partial_r\Psi-\Psi\partial_r\Psi^*).
\ee
Following \cite{Gonzalez_2021} we can solve the Klein-Gordon equation in three different regions matching the solutions to produce the amplification factor of real frequency waves $\omega$, as
\be
\label{ee3}
\mathcal{R} = \frac{|\omega_1|^2r_h^3 + r_h + 2r_h^2\omega_1 }{|\omega_1|^2r_h^3 + r_h - 2r_h^2\omega_1  }
\ee
in which $\omega_1=\omega -\Phi_h$. We find the same turning point for the amplification factor as of (\ref{e9}), $\omega_1 >0$ which results in $\mathcal{R}>1$. The instabilities of the scalar field we found are robust and physically motivated as we can see from the above calculations.

\section{Discussion}\label{sec5}

In this paper we studied a charged scalar field propagating in the nonspinning  charged BTZ black hole. We verify that small charge fields have stable evolutions in the fixed geometry, providing a quasinormal spectrum that depends on the geometry parameters (i. e. black hole charge, mass and AdS radius).

With two different numerical methods we calculate those quasinormal spectra examining the influence of the scalar/geometric charges in it verifying that the black hole charge acts in different directions for the real and imaginary parts of $\omega$. In this sense, the key aspect to understand the quasinormal spectrum is the presence of a particular scalar charge $q_c$ as of the scalar field destabilizes the geometry. 

Whenever the propagating field has $q<q_c$, its evolution is dictated by a quasinormal spectrum whose fundamental mode diminishes (both imaginary and real parts) as $\mathcal{Q}$ increases. For high scalar charges though $q \gg q_c$, not only the field profile is unstable, but the instability increases with the augmentation of $\mathcal{Q}$. Similarly, in such regime, $\Re (\omega )$ increases with $\mathcal{Q}$.

The field instabilities are also unveiled through the potential analysis as demonstrated in figures \ref{fig1}: the higher the potential depth, the more likely the scalar field is to evolve unstably. In that sense, for every geometry configuration scalar fields with sufficient high charges are unstable. 

The shape of the electromagnetic potential produces interesting nontrivial consequences for the field evolution. Since it is represented by a monotonically increasing function, the higher the scalar charge the more unstable the field profile evolves. This is a specific behavior of the BTZ black holes not appearing in multiple different charged spacetimes \cite{oli21, kono1, Zhu_2014}. As a consequence, the typical timescale of the threshold region of stability changes drastically both at instability and oscillatory levels if we consider higher scalar charges. 

The expected timescales are such that $\tau_i \gg \tau_o \gg \tau_{BH} (\sim r_h)$. This seems to be true for small scalar charges until a near value $q \sim q_c$ from which such approximation is not any more valid. In this range we calculate the energy flux across the event horizon and the amplification factor of real wave scatterings finding a similar condition for superradiance in both cases.

Possible lines of investigation in the topic include the study of other probe fields in the same geometry and in the rotating black hole with charge.

\section{Acknowledgments}

The author thanks the hospitality of DMAT - University of Aveiro/CIDMA. He is grateful to Carlos A. R. Herdeiro, Clovis A. S. Maia and Jefferson S. E. Portela for fruitful discussions and João E. K. Nauderer for help in running the numerical recipes. The work is supported by the Center for Research and Development in Mathematics and Applications (CIDMA) through the Portuguese Foundation for Science and Technology (FCT— Fundação para a Ciência e a Tecnologia), references https://doi.org/10.54499/UIDB/04106/2020 and https:// doi.org/10.54499/UIDP/04106/2020.

\appendix

\section{Instability conditions to the Klein-Gordon charged field}\label{app1}

\begin{figure}
\begin{center}
\includegraphics[width=0.6\textwidth]{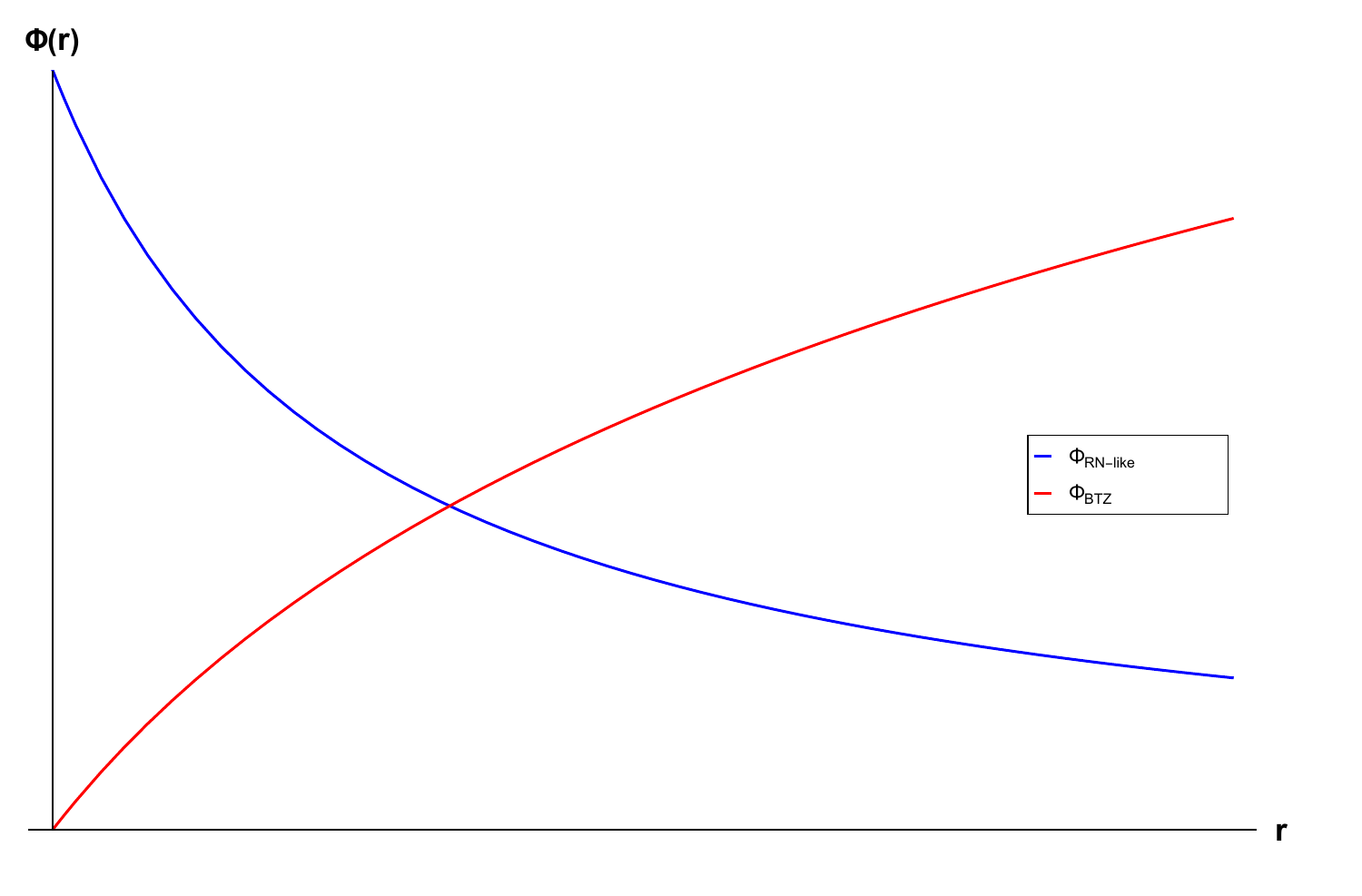}
\end{center}
\caption{The electromagnetic potential of charged spacetimes: BTZ geometry and RN-like black holes (pure, dS, AdS, with fluids). Equation (\ref{ap3}) is satisfied when $\omega_r$ lies within the scope of $\Phi$, what in the BTZ case brings inequality (\ref{ap5}) as a result.}
\label{fig3}
\end{figure}

In order to obtain the first condition of section \ref{sec3}, we start by considering the scalar equation in a similar form as in (\ref{e5}),
\be
\label{ap1}
\psi'' + (\omega - \Phi )^2 \psi - V(r) \psi =0
\ee
in which $\Phi = -qA_t$ is the electric potential, $\psi (t) \rightarrow e^{-i \omega t }$ (the usual temporal Ansatz for the field was used)  and prime denotes derivation relative to $r_*$. Mutiplying (\ref{ap1}) by $\psi^*$, substituting $\psi^* \psi'' = (\psi^* \psi ')' -|\psi '|^2$ and integrating the remaining equation we obtain 
\be
\label{ap2}
(\psi^*\psi') \Big|_{-\infty}^{0} + \int_{-\infty}^{0} (\omega - \Phi )^2 |\psi|^2dr_* =  \int_{-\infty}^{0} (|\psi '|^2 + V(r) |\psi|^2 )dr_* 
\ee
Now, the rhs of equation (\ref{ap2}) is real. If we consider the imaginary part of (\ref{ap2}), we have
\be
\label{ap3}
{\Im} (\psi^*\psi') \Big|_{-\infty}^{0}  + 2\int_{-\infty}^{0} (\omega_i \omega_r - \omega_i\Phi) |\psi|^2dr_* = 0
\ee
where for simplicity we rewrite $\Im (\omega ) = \omega_i$ and  $\Re (\omega ) = \omega_r$. Now, if $\omega_i >0$, 
\be
\label{ap4}
(\psi^*\psi') \Big|_{-\infty}^{0} = (\psi^*\psi') \Big|^{0} - (\psi^*\psi') \Big|_{-\infty} =  i e^{2\omega_i r_*}(\omega - \Phi_h) \Big|_{-\infty} =0
\ee
Since $\Phi$ is a monotonically increasing function between $r_h$ and $r_\infty$ - a very specific condition of the charged BTZ spacetime and different from the 4-dimensional case (see figure \ref{fig3}) - it has its minimum value at $r_h$, $\omega_r$ must lie between both limits of $\Phi$ so that (\ref{ap3}) holds. Then
\be
\label{ap5}
\omega_r > \Phi_h
\ee
is the necessary, but not sufficient condition for the existence of unstable modes. This inequality is similar to the designated "superradiant" condition for quasinormal modes as that found in \cite{kono1}.

The second condition for instability or the destabilization of the scalar field comes from the effective potential. 
Let us consider equation (\ref{ap1}) in a different form,
\be
\label{ap6}
\psi'' - \ddot\psi - U \psi =0
\ee
in which dot is the derivative with relation to $t$, prime to $r_*$ and $U = V(r) -\Phi^2 - 2\omega \Phi$. Let us apply $\psi = e^{-i\omega r_*}\Psi$ in (\ref{ap6}), multiply the resultant equation by $\Psi^*$, substitute $\Psi^* (f\Psi')'= (f\Psi^* \Psi ')' - f|\Psi'|^2$ and integrate between $r_h$ and AdS infinity. Then equation (\ref{ap6}) turns to
\be
\nonumber
-\int_{r_h}^{\infty}f|\Psi'|^2dr - 2i\omega \int_{r_h}^{\infty}\Psi^*\Psi' dr - 2\int_{r_h}^{\infty}f^{-1}\omega q A_t |\Psi|^2dr - \int_{r_h}^{\infty}f^{-1}\Big( V(r) + q^2 A_t^2 \Big)|\Psi|^2dr =0\\
\label{ap7}
\ee 
in which we take $f\Psi^*\Psi'\Big|_{r_h}^{\infty}=0$. We will replace the second and third terms of this equation with a series of operations in what follows. Let us consider the imaginary part of (\ref{ap7}),
\be
\label{ap8}
\Im \Big( - 2i\omega \int_{r_h}^{\infty}\Psi^*\Psi' dr \Big) + 2\omega_i q  \int_{r_h}^{\infty}A_t f^{-1}|\Psi|^2 dr =0.
\ee
Now taking the first term of the above equation ($\equiv \bar{\Im}$) as
\be
\nonumber
\bar{\Im} = - \int_{r_h}^{\infty}\omega_r(\Psi^* \Psi' +{\Psi^*}' \Psi )dr - i\omega_i 
\int_{r_h}^{\infty}(\Psi^* \Psi' -{\Psi^*}' \Psi )dr = \\
\nonumber
 \frac{\omega_r}{\omega}\Bigg(- \int_{r_h}^{\infty}\omega(\Psi^* \Psi)'dr +2 i\omega_i 
\int_{r_h}^{\infty}{\Psi^*}' \Psi dr \Bigg) =  \frac{\omega_r \omega_i}{|\omega|^2} \Bigg(\frac{|\omega|^2 |\Psi|^2}{\omega_i}\Bigg|_{r_h}^{\infty} - 2i\omega \int_{r_h}^{\infty}\Psi ' \Psi^* dr \Bigg)^* \\
\label{ap9}
\ee
and substituting eqs. (\ref{ap9}) and (\ref{ap8}) into (\ref{ap7}) we get
\be
\label{ap10}
\int_{r_h}^{\infty}\Bigg(f|\Psi'|^2 + f^{-1}(V(r)-q^2A_t^2)|\Psi|^2 \Bigg)dr = -\frac{|\omega|^2|\Psi|^2}{\omega_i} \Bigg|_{r_h}^{\infty}.
\ee
In this relation we can see the necessary (but not sufficient) condition for the field destabilization, $\omega_i >0$, that of equation (\ref{e10}).




\bibliography{references}

\end{document}